\documentclass[12pt,preprint]{aastex}

\def\ovii{\ion{O}{7}}
\def\oviii{\ion{O}{8}}
\def\neix{\ion{Ne}{9}}

\def\kms{$\:\rm{\,km\,s}^{-1}$}
\def\n_et{$n_{\rm e}t$}
\def\cm3s{$\rm{\,cm}^{-3}\,$s}

\shorttitle{Dynamics of X-Ray--Emitting Ejecta in Puppis~A} 
\shortauthors{Katsuda et al.}

\begin{document}

\title{Dynamics of X-Ray--Emitting Ejecta in the Oxygen-Rich Supernova 
Remnant Puppis~A Revealed by the {\it XMM-Newton} RGS}

\author{Satoru Katsuda\altaffilmark{1}, Yutaka Ohira\altaffilmark{2},
Koji Mori\altaffilmark{3}, Hiroshi Tsunemi\altaffilmark{4}, 
Hiroyuki Uchida\altaffilmark{5}, \\
Katsuji Koyama\altaffilmark{4, 5}, and Toru Tamagawa\altaffilmark{1}
}

\altaffiltext{1}{RIKEN (The Institute of Physical and Chemical
Research), 2-1 Hirosawa, Wako, Saitama 351-0198}

\altaffiltext{2}{Department of Physics and Mathematics, Aoyama Gakuin 
University, 5-10-1 Fuchinobe Sagamihara, 252-5258 Japan}

\altaffiltext{3}{Department of Applied Physics, Faculty of Engineering,
University of Miyazaki, 1-1 Gakuen Kibana-dai Nishi, Miyazaki, 889-2192,
Japan}

\altaffiltext{4}{Department of Earth and Space Science, Graduate School
of Science, Osaka University, 1-1 Machikaneyama, Toyonaka, Osaka,
60-0043, Japan}

\altaffiltext{5}{Department of Physics, Kyoto University, 
Kitashirakawa-oiwake-cho, Sakyo, Kyoto 606-8502, Japan}





\begin{abstract}
Using the unprecedented spectral resolution of the reflection grating 
spectrometer (RGS) onboard {\it XMM-Newton}, we reveal dynamics of 
X-ray--emitting ejecta in the oxygen-rich supernova remnant Puppis~A.  
The RGS spectrum shows prominent K-shell lines, including \ovii\ 
He$\alpha$ forbidden and resonance, \oviii\ Ly$\alpha$, \oviii\ Ly$\beta$, 
and \neix\ He$\alpha$ resonance, from an ejecta knot positionally 
coincident with an optical oxygen-rich filament (the so-called 
``$\Omega$" filament) in the northeast of the remnant.  We find that 
the line centroids are blueshifted by 1480$\pm$140$\pm$60\kms~(the first 
and second term errors are measurement and calibration uncertainties, 
respectively), 
which is fully consistent with that of the optical $\Omega$ filament.  
Line broadening at 654\,eV (corresponding to \oviii\ Ly$\alpha$) is 
obtained to be $\sigma\lesssim0.9$\,eV, indicating an oxygen temperature 
of $\lesssim30$\,keV.  Analysis of {\it XMM-Newton} MOS spectra shows 
an electron temperature of $\sim$0.8\,keV and an ionization timescale 
of $\sim$2$\times$10$^{10}$\,\cm3s.  We show that the oxygen and electron 
temperatures as well as the ionization timescale can be reconciled if 
the ejecta knot was heated by a collisionless shock whose velocity is 
$\sim$600--1200\kms\ and was subsequently equilibrated due to Coulomb 
interactions.  The RGS spectrum also shows relatively weak K-shell lines 
of another ejecta feature located near the northeastern edge of the 
remnant, from which we measure redward Doppler velocities of 
650$\pm$70$\pm$60\kms.
\end{abstract}
\keywords{ISM: abundances --- ISM: individual objects: Puppis~A --- 
ISM: supernova remnants --- X-rays: ISM}

\section{Introduction}

The Galactic supernova remnant (SNR) Puppis~A is one of the 
so-called ``oxygen-rich" SNRs, in which oxygen-rich fast-moving optically 
emitting ejecta knots (OFMKs) are identified.  So far, eight other SNRs, 
Cas~A and G292.0+1.8 in our Galaxy and six others in the Magellanic 
Clouds and NGC~4449, are categorized in this class 
\citep[][for a recent summary]{Vink2012}.  Of these, Puppis~A is 
particularly interesting, because all of OFMKs are located in the eastern and 
northeastern (NE) portion \citep[][]{Winkler1985,Winkler1988}, 
while the stellar remnant is moving in the opposite southwestern direction 
\citep{Petre1996,Hui2006,Winkler2007,Becker2012}.  This is consistent 
with recent models, which predicts that SN ejecta are expelled 
preferentially in one direction, while the stellar remnant receives a kick 
in the opposite direction \citep[][and references therein]{Wongwathanarat2012}.
Thus, Puppis~A is an extremely important target for the study of SN 
explosion mechanisms as well as SN nucleosynthesis.

Puppis~A is one of the brightest SNRs in the X-ray sky.  The X-ray 
emission has been thought to be dominated by strong cloud-shock 
interactions \cite[][references therein]{Charles1978,Petre1982}.  
A large area of the remnant 
was recently mapped by X-ray observatories in orbit, i.e., 
{\it XMM-Newton}, {\it Chandra}, and {\it Suzaku}.  They confirmed 
earlier results that the X-ray emission is dominated by the swept-up 
interstellar medium \citep[][]{Hwang2005,Hwang2008,Katsuda2010}.  
On the other hand, the data also revealed signatures of SN ejecta in 
three locations: 
a Si-rich clump \citep{Hwang2008}, a bright O-Ne-Mg-(Si-S)-rich knot 
\citep[the ejecta knot called hereafter:][]{Katsuda2008}, whose 
southern portion is positionally coincident with an ``$\Omega$"-shaped 
OFMK \citep[the so-called $\Omega$ filament named by][]{Winkler1985}, 
and an O-Ne-Mg-rich filament \citep[the ejecta filament called 
hereafter:][]{Katsuda2010}.  All of them are located in the 
NE quadrant, further supporting the one-sided ejection of SN debris.

While picking up ejecta features reveal two dimensional ejecta 
distributions on the plane of the sky, measurements of Doppler 
velocities can provide us with line-of-sight structures \citep[][for a 
recent review of kinematics of X-ray SNRs]{Dewey2010}.  \citet{Katsuda2008} 
found that the ejecta knot described above shows blueshifted 
K-shell lines.  The Doppler velocities were measured to be 
3400$^{+1000}_{-800}$\kms\ at the northern knot and 
1700$^{+700}_{-800}$\kms\ at the southern knot which agrees with the 
optical measurement of $\sim$1500\kms\ \citep{Winkler1985}.  
The possible different Doppler velocities between the north and the south 
led us to suggest that the ejecta knot actually consists of two different 
knots which appear to be located close with each other.  However, the 
poor quality of the data prevented us from making strong arguments; the 
X-ray data were taken by the {\it XMM-Newton} European Photon Imaging 
Camera \citep[EPIC:][]{Turner2001,Struder2001} with moderate spectral 
resolution ($E/\Delta E\sim20$), and also the ejecta knot was detected 
at the edge of the field of view (FOV) where calibration uncertainties 
would be large.

Here, we report on precise Doppler measurements for the ejecta knot and 
the ejecta filament in Puppis~A, based on the unprecedented spectral 
resolution X-ray spectra ($E/\Delta E\sim150$) obtained by the 
Reflection Grating Spectrometer 
\citep[RGS:][]{denHerder2001} onboard {\it XMM-Newton}.  
In addition, we give a restrictive upper limit on an oxygen temperature of 
the ejecta knot, for the first time.  These new measurements allow us to 
discuss the nature of the ejecta knot in detail.

\section{Observation and Data Reduction}

\begin{figure}
\includegraphics[angle=0,scale=0.5]{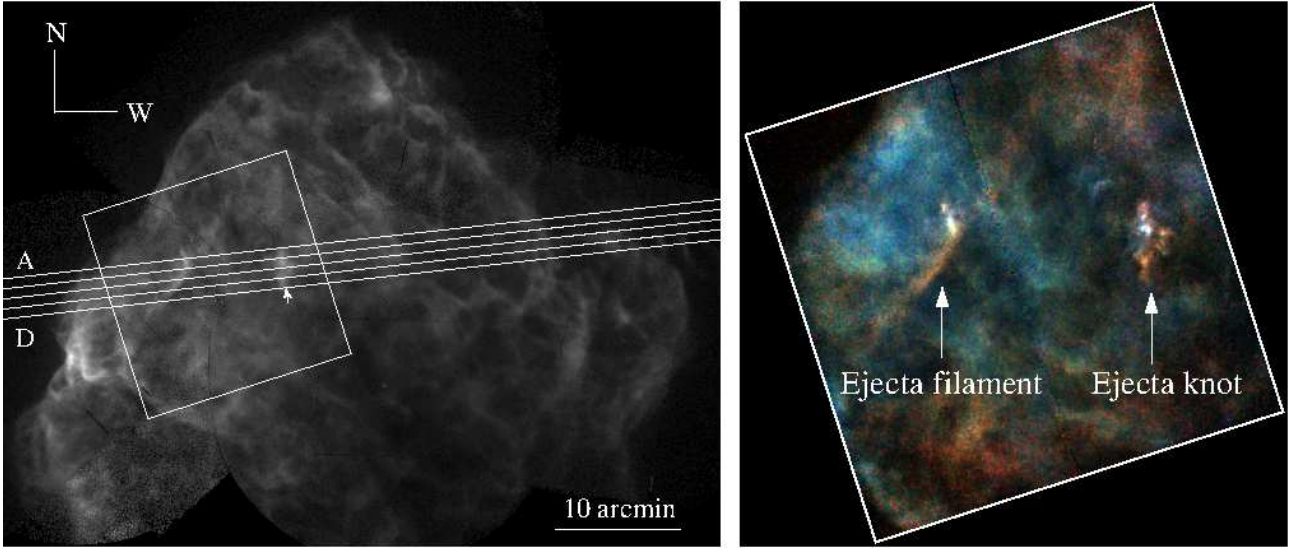}\hspace{1cm}
\caption{Left: The RGS spectral extraction regions overlaid on 
  the merged {\it XMM-Newton} and {\it Chandra} image in the energy band 
  0.5--5\,keV.   The arrow indicates zero points of 
  projection profiles shown in Figure~\ref{fig:prof}.  
  Right: Three-color close-up {\it Chandra} image of the white box region 
  in the left panel.  Red, green, and blue correspond to 0.5--0.7\,keV, 
  0.7--1.2\,keV, and 1.2--5.0\,keV, respectively.  Two ejecta features 
  focused in this paper are indicated as arrows.
} 
\label{fig:image}
\end{figure}

We performed an {\it XMM-Newton} observation of Puppis~A on 2012 October 
20 (Obs.ID 0690700201), aiming at the fast-moving ejecta knot coincident
with the optical $\Omega$ filament.  Our primary instrument in this 
observation is the RGS that can produce high-resolution X-ray spectra, 
while we also make use of the EPIC data to support the RGS results as 
well as to obtain additional information.  The dispersion direction 
of the RGS is 96$^{\circ}$.4 measured east to north, as we show 
in Figure~\ref{fig:image} left---an X-ray image of Puppis~A 
generated by existing {\it XMM-Newton} and {\it Chandra} data.  
Fortunately, no background flare is seen in our light curve, allowing us 
to use full exposure times of 20.8\,ks, 12.1\,ks, and 11.7\,ks for the 
RGS1/2, MOS1, and MOS2, respectively.  All the raw data were processed 
using version 11.0.0 of the XMM Science Analysis Software and the 
calibration data files available in 2012 May.

\section{RGS Spectra}

The slitless RGS spectrometer is generally not useful for extended 
sources like Puppis~A, because off-axis emission along the dispersion 
direction is detected at wavelength positions shifted with respect 
to the on-axis source, which causes degradation of energy resolution.  
However, if the angular size of the target is small enough ($\lesssim$ a 
few arc minutes) and is brighter than its surroundings, it is possible to 
obtain high-resolution spectra for them.  Our on-axis source, 
the ejecta knot, is such a lucky case, for which we are able to obtain an 
order-of-magnitude better resolution spectra than nondispersive CCDs.  
There is another ejecta feature, i.e., the ejecta filament described in 
the introduction, within the RGS FOV.  Since the angular size 
of the filament along the dispersion direction is as thin as 1$^{\prime}$, 
we can also find its signatures in the RGS spectrum.  
The two ejecta features of interest are indicated on a {\it Chandra} 
image shown in Figure~\ref{fig:image} right.

The analysis method basically follows our previous work \citep{Katsuda2012}.  
Briefly, we first extract four RGS spectra from 0$^\prime$.8-width 
sectors along the cross-dispersion direction.  We then smooth the RGS 
response designed for on-axis point sources, based on emission 
profiles along the dispersion direction by using the {\tt rgsrmfsmooth} 
software.  Finally, we fit the RGS spectrum with emission models in the 
XSPEC package \citep{Arnaud1996}.  
The RGS background (BG) is generated from a blank-sky observation 
(Lockman hole: Obs.ID 0147511601), as we did in our previous work.  
The blank-sky BG emission is less than 1\% of the source emission in 
the energy band analyzed.  

\begin{figure}
\includegraphics[angle=0,scale=0.65]{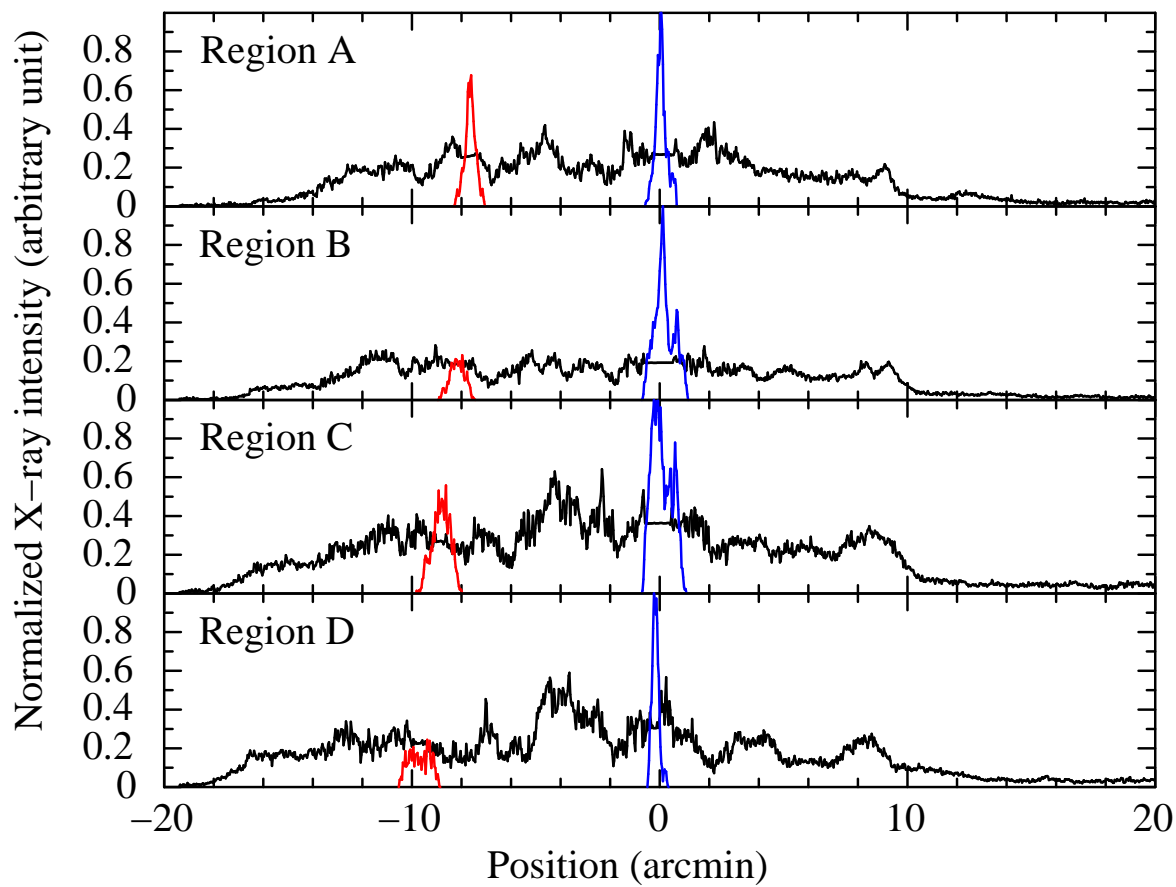}\hspace{1cm}
\caption{X-ray emission profiles in 0.6--0.7\,keV along the RGS 
  dispersion axis for the four spectral extraction regions.  The three 
  profiles in each panel are responsible for the ejecta knot in blue, 
  the ejecta filament in red, and the diffuse surroundings in black.
  The X-ray intensities are normalized at the peak values of the 
  ejecta knot, showing how diffuse surrounding emission is significant
  in the RGS spectra.  The zero points in x-axis (x=0) correspond to 
  the direction from which emission is detected at the nominal (no 
  red/blue shift) wavelength positions on the RGS detectors.  
} 
\label{fig:prof}
\end{figure}

Here, the RGS response in each region is smoothed with three distinct 
emission profiles responsible for (1) the ejecta knot, (2) the ejecta 
filament, and (3) their surrounding diffuse emission, so that we can 
derive physical parameters for individual features.  Emission profiles 
are generated with either the {\it Chandra} or {\it XMM-Newton} image, 
for which we take account of vignetting effects of {\it XMM-Newton}'s 
X-ray telescope and the asymmetry of the effective area along the 
dispersion axis described in the literature \citep[][]{Broersen2013}.
The emission profiles for the ejecta features are extracted from 
{\it Chandra} images masked for them, for which we subtract 
underlying diffuse emission by interpolating the surrounding diffuse 
emission.  We have checked that, even if we change the intensity of 
underlying diffuse emission by $\pm$20\% which is a small-scale 
inhomogeneity around the ejecta knot, the fit results presented below 
are not affected significantly.
Figure~\ref{fig:prof} shows the X-ray emission profiles in 
0.6--0.7\,keV for the four regions.  The three profiles in each panel 
are responsible for (1) the ejecta knot in blue, (2) the ejecta filament 
in red, and (3) the surroundings in black.

\begin{figure}
\vspace{-1cm}
\includegraphics[angle=0,scale=0.6]{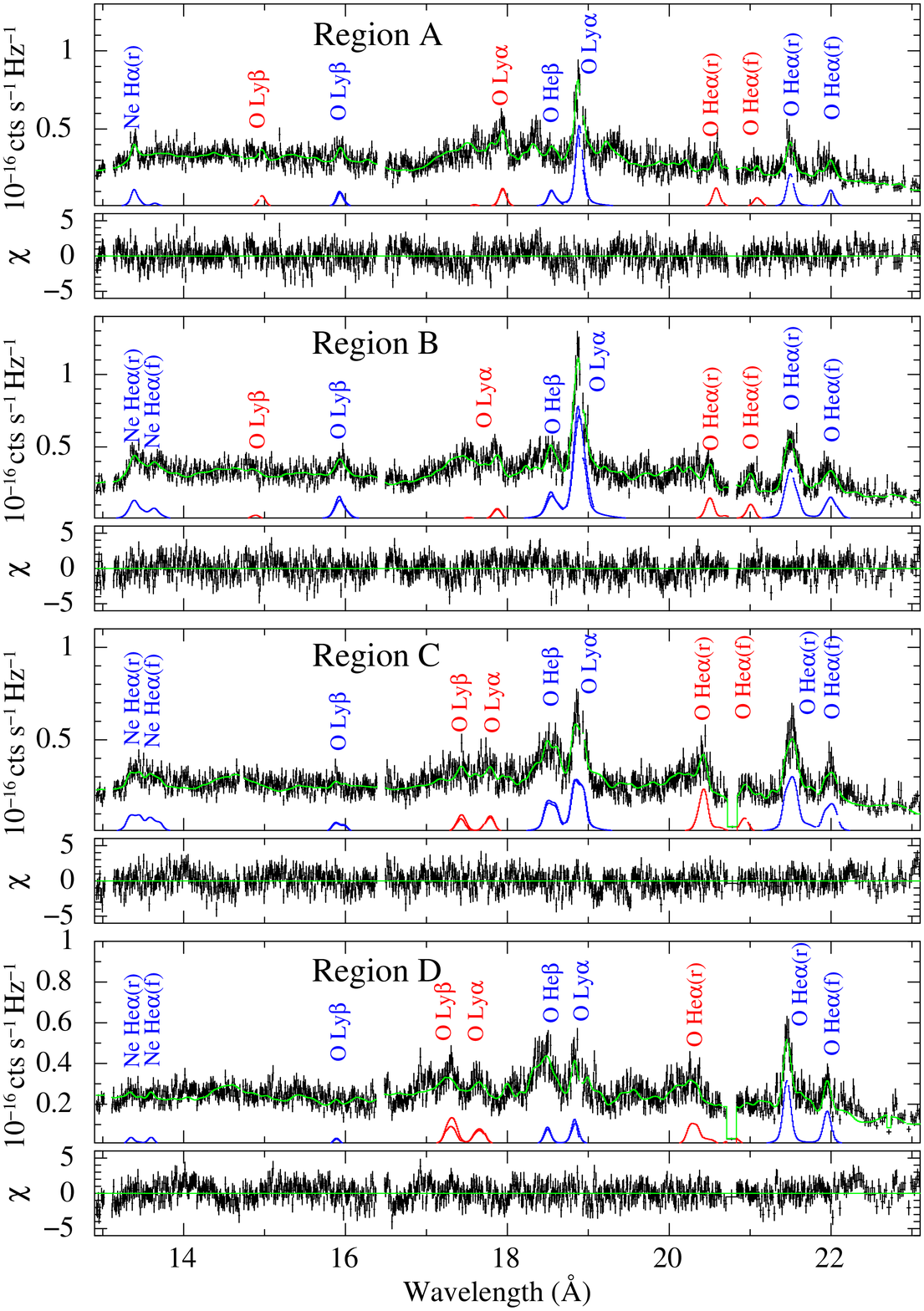}\hspace{1cm}
\vspace{-0.5cm}
\caption{{\it XMM-Newton} spectra of the four regions in 
  Figure~\ref{fig:image}.  The RGS1 and RGS2 cover 15.5--24\,\AA\ 
  ($\sim$0.8--0.51\,keV) and 13--19.1\,\AA\ ($\sim$1--0.65\,keV), 
  respectively.  The two (RGS1 and RGS2) data sets are simultaneously 
  fitted with a combination model of {\tt vpshock} (for diffuse background 
  emission) + nine Gaussians (for the ejecta knot) + six Gaussians 
  (for the ejecta filament).  The total best-fit model is shown in green, 
  while individual components are in blue and red for the ejecta knot and 
  the ejecta filament, respectively.  Lower panels show the residuals.
} 
\label{fig:RGSspec1}
\end{figure}

\begin{figure}
\vspace{-1cm}
\includegraphics[angle=0,scale=0.65]{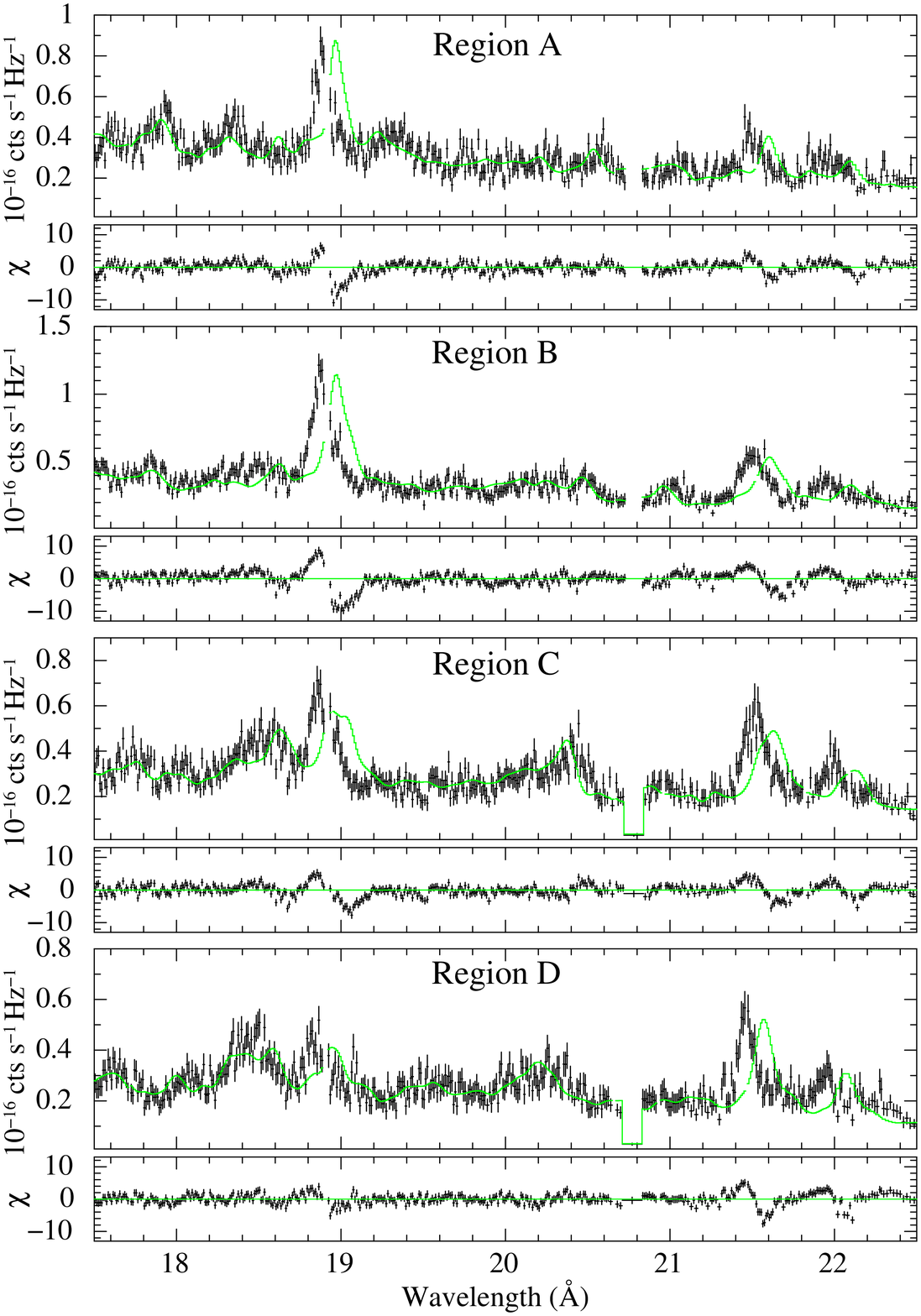}\hspace{1cm}
\vspace{-1cm}
\caption{Close-up RGS spectra for O K-shell lines.  To clarify the 
  Doppler shifts of the two ejecta features, we here show the best-fit 
  models shifted at rest-frame positions, i.e., redshift = 0.  
  The lower panels show residuals.  Blueshifts for O Ly$\alpha$ and 
  O He$\alpha$ for the ejecta knot component are clearly seen.  
} 
\label{fig:RGSspec2}
\end{figure}

Figures~\ref{fig:RGSspec1} and \ref{fig:RGSspec2} shows BG-subtracted 
RGS1 and RGS2 spectra for the four regions.  While most of the emission 
is due to overall diffuse emission which acts as effective BG emission 
in our analysis, several peaks can be identified as line emission from 
our interesting two ejecta features.  Some of them are indicated in the 
plot as blue and red labels for the ejecta knot and the ejecta filament, 
respectively.  
Note that there is large wavelength shifts in line centroids between 
the knot (blue) and the filament (red) by about 1\AA.  This is mainly 
due to the off-axis position (about 8$^{\prime}$) of the filament; an 
offset of 1$^{\prime}$ causes a wavelength shift of 0.138\,\AA.  
We use data in the energy range of 0.5--1\,keV 
(24--13\,\AA) where line emission from the ejecta features are clearly 
detected.  Dividing the energy band into four sections, 0.5--0.6\,keV, 
0.6--0.7\,keV, 0.7--0.85\,keV, and 0.85--0.97\,keV, we generate RGS 
responses using the same energy-band images.  We note that the available 
energy ranges for the RGS1 and the RGS2 are limited to 0.5--0.8 
(24--15.5\,\AA) and 0.65--1\,keV (19--13\,\AA), respectively, since 
chip \#7 in the RGS1 and chip \#4 in the RGS2 are no longer operating.

The RGS spectra consist of diffuse emission and emission from 
the ejecta knot and ejecta filament.  We model the diffuse emission and 
the ejecta in terms of a nonequilibrium ionization shock model 
\citep[the {\tt vpshock} model:][]{Borkowski2001} and individual 
Gaussian components, respectively.  We use nine gaussian components for 
the ejecta knot and six for the ejecta filament.  We impose the 
Gaussians in the knot component to be 
\ovii~He$\alpha$ forbidden at @0.561\,keV (22.1\,\AA), 
\ovii~He$\alpha$ intercombination at @0.568\,keV (21.8\,\AA), 
\ovii~He$\alpha$ resonance at @0.574\,keV (21.6\,\AA), 
\oviii~Ly$\alpha$ @0.654\,keV (19.0\,\AA), 
\ovii~He$\beta$ @0.666\,keV (18.6\,\AA), 
\oviii~Ly$\beta$ @0.775\,keV (16.0\,\AA), 
\neix~He$\alpha$ forbidden @0.905\,keV (13.7\,\AA), 
\neix~He$\alpha$ intercombination @0.915\,keV (13.6\,\AA), and
\neix~He$\alpha$ resonance @0.922\,keV (13.5\,\AA).  
The six Gaussians for the filament component are the same O K-shell 
lines listed above.  For the diffuse component, the degradation of RGS 
spectra is sever owing to the large spatial extent; the effective 
spectral resolution is much worse than the nondispersive CCDs.  Therefor, 
it is difficult to constrain numerous parameters, hence in our fitting 
procedure, we only allow the electron temperature, $kT_{\rm e}$, and 
normalizations to vary freely, fixing the intervening hydrogen column 
density, $N_\mathrm H$, to be 3$\times$10$^{21}$\,cm$^{-2}$ 
and the range of the ionization timescale, \n_et, from zero up to 
3$\times$10$^{11}$\,cm$^{-3}$\,s, based on recent X-ray measurements 
\citep{Hwang2005,Hwang2008,Katsuda2010}.  
For the Gaussian components, we allow all
normalizations to vary freely.  Line centroids are fixed to theoretically 
expected values \citep{Smith2001}, but we allow a redshift parameter 
to vary freely, keeping the same value for all Gaussians in 
either the knot or the filament component.  Line broadening of the 
knot component is a free parameter taken to be proportional to line center 
energies as is expected for a collisionless shock heating.  That of the 
filament component is set to zero, since line emission from the filament 
is not very clear in the RGS spectrum.  For the knot component in regions 
A, B, and C, we measure redshift and line broadening separately for the 
RGS1 and the RGS2, so that we can evaluate systematic uncertainties in 
our measurements and also see if energy dependence is present; the RGS1 
covers 0.5--0.8\,keV including O K-shell lines while the RGS2 covers 
0.65--1\,keV including O and Ne K-shell lines, respectively.  
Before BG subtraction, each spectrum is grouped into bins 
with at least $\sim$25 counts, which allows us to perform a $\chi^2$ test.


With this fitting strategy, we obtain fairly good fits for all the 
spectra as shown in Figure~\ref{fig:RGSspec1}, where individual knot 
and filament components are shown in blue and red, respectively, as 
well as the total model in green.  
The fit results are summarized in Table~\ref{tab:RGSresults}.
Overall, the redshifts and line broadening derived from the RGS1 and 
the RGS2 are consistent with each other, suggesting no significant 
energy dependence for these parameters.  We also fitted the 
RGS2 spectra, by using a specific energy band for \neix\ He$\alpha$ 
lines, finding consistent results as in Table~\ref{tab:RGSresults}.
Figure~\ref{fig:RGSspec2} shows close-up RGS spectra for O K-shell 
lines.  In this plot, we show the best-fit models shifted at 
rest-frame positions (i.e., redshift = 0) in green.  The Doppler 
shifts of the ejecta knot are clearly seen.

The blueshift measured at the southernmost region (region D), 
1560$\pm$60\kms, is consistent with our previous X-ray measurements 
\citep{Katsuda2008}, significantly reducing the uncertainty.
It also matches with the previous optical measurements for the $\Omega$ 
filament: 1520$\pm$40\,\kms\ or 1570$\pm$20\,\kms\ from the SIT or CCD 
spectrum, respectively \citep{Winkler1985}.  While the blueshifts 
in the four regions are all close with each other, that in the 
northernmost region is slightly slower than the other regions.  
This conflicts with our previous measurements that suggested 
a faster velocity in the north of the knot than that in the south.  
Given much improved spectral resolution in the current data, we 
believe that the result presented here is more reliable than 
the previous one.  As for the slightly slower velocity, it may indicate 
that the northern part is responsible for a tail structure delayed from 
the main body of the ejecta knot.  We find, for the first time, that 
line emission from ejecta filament is redshifted by $\sim$650\kms.


Line broadening in the ejecta knot is found to be fairly narrow.  
It ranges from 0 to 0.9\,eV at 0.654\,keV in the four regions, which we 
take as our measurement uncertainty.  Based on the relation between 
temperatures and line broadening, $\sigma = E_0 \sqrt{kT/mc^2}$ 
\citep{Rybicki1979}, we obtain an upper limit of an oxygen 
temperature to be 30\,keV.

Line intensity ratios measured are consistent with those expected by 
a non-equilibrium ionization plasma model.  We have indeed confirmed 
that the {\tt vnei} model \citep{Borkowski2001} can also well reproduce 
the RGS spectra (values of $\chi^2$/d.o.f.\ range from 1.4 to 1.6) with 
plausible electron temperatures, $\sim$0.4--1\,keV, an ionization 
timescale of 2$\times$10$^{10}$\cm3s, and both Doppler shifts and line 
broadening consistent with those in Table~\ref{tab:RGSresults}.

\section{MOS Spectra}

\begin{figure}
\includegraphics[angle=0,scale=0.4]{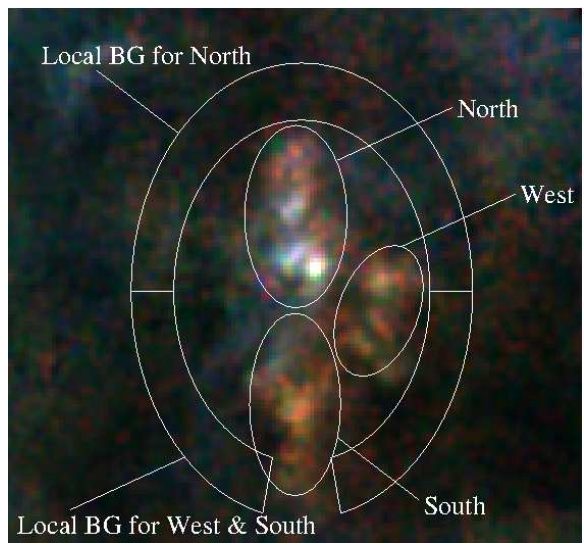}\hspace{1cm}
\caption{Spectral extraction regions for the MOS data 
  overlaid on a close-up {\it Chandra} image shown in Figure~\ref{fig:image} 
  right.  The local BG regions are also shown.
} 
\label{fig:MOSregions}
\end{figure}

\begin{figure}
\includegraphics[angle=0,scale=0.5]{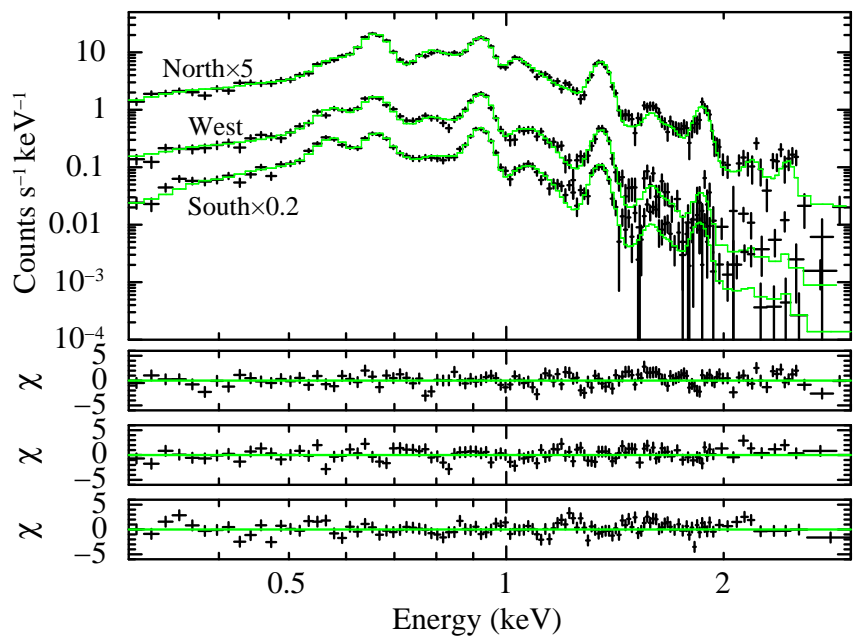}
\caption{The MOS spectra extracted from the regions shown in 
  Figure~\ref{fig:MOSregions}.  The best-fit {\tt vnei} models are 
  shown together in green.  For clarity, the spectra from the north 
  and the south are multiplied by 5 and 0.2 times, respectively.  
  The lower panels are the residuals for north, west, and south 
  from top to bottom, respectively.
} 
\label{fig:MOSspec}
\end{figure}

Since the RGS spectra are heavily contaminated by the surrounding 
diffuse emission in our case, electron temperatures and ionization 
timescales for our interesting features 
are better constrained by the MOS rather than the RGS.  
We thus analyze the MOS spectra for the ejecta knot.  As shown in a 
close-up {\it Chandra} image in Figure~\ref{fig:MOSregions}, the knot 
appears to consist of three parts: north, west, and south.  We 
extract spectra from these features.  Local BGs are taken from the 
adjacent regions.  We sum up data taken by the MOS1 and the MOS2 to 
improve photon statistics.  The local-BG subtracted MOS1+2 spectra 
together with the best-fit {\tt vnei} model 
\citep[the augmented version 2.0:][]{Borkowski2001}, 
are presented in Figure~\ref{fig:MOSspec}.

It is difficult to measure absolute (relative to hydrogen) metal 
abundances for metal-rich plasmas, because continuum emission comes 
from the metals themselves and no longer reflects the amount of hydrogen, 
and also because in our case the continuum emission from the ejecta knot 
is difficult to estimate adequately, given that a considerable 
fraction of the X-ray emission is due to local BG that cannot be 
subtracted perfectly.  Therefore, in our spectral modeling, we examine 
two abundance patterns: In case A, we assume the Fe/H abundance to be 
the solar value \citep{Anders1989} as line emission from Fe is not evident 
in our X-ray spectra, while in case B, we assume the O/H abundance to be 
2000 times the solar value which is derived from optical spectroscopy 
of the $\Omega$ filament \citep{Winkler1985}.  
We find that the {\tt vnei} model represents the 
data fairly well for both of the two cases.  The best-fit models presented 
in Figure~\ref{fig:MOSspec} are those for case A which gives slightly better 
fits than case B.   Table~\ref{tab:MOSresults} summarizes the results and 
notes details of our fitting. 

The blueshifts from the MOS spectra are in reasonable agreements with 
the RGS measurements, considering the relatively large calibration 
uncertainty ($\pm$5\,eV\footnote{http://xmm2.esac.esa.int/docs/documents/CAL-TN-0018.ps.gz}) on the energy scale of the MOS.
We also confirm the elevated abundances of O, Ne, and Mg compared with Fe.  
On the other hand, the electron temperatures and ionization timescales 
are outside statistical uncertainties of our previous measurements 
\citep{Katsuda2008}.  The discrepancy would be partly due to the different 
data quality; the data were taken at the edge (previous) or center (this 
time) of the FOV, and partly due to the different fitting strategy; we here 
use (1) the most recent version of the {\tt vnei} code, (2) different spectral 
extraction regions, and (3) different energy band, and also we fix 
$N_{\mathrm H}$ to be 3$\times$10$^{21}$\,cm$^{-2}$ \citep{Katsuda2010}, 
but allow abundances of C and N to vary freely.  Not only these analysis 
improvements but also the fact that blueshifts derived here are closer to the 
RGS measurements assure reliability of the current results rather than 
the previous ones.


Assuming the plasma depths to be 1$^{\prime}$ \citep[or $\sim$0.6\,pc at 
a distance of 2.2\,kpc:][]{Reynoso2003}, which is roughly the size of 
the knot, we calculate densities and masses in each region.  Although it 
is often assumed that a density ratio between electrons and protons to 
be unity or so, this assumption is incorrect for super-metal rich plasmas 
like the ejecta knot in Puppis~A.  Therefore, we here calculate the 
ratio, taking account of abundances as well as ion fractions based on 
the SPEX code \citep{Kaastra1996}.  Then, the emission measures and 
abundances give densities and masses for each species as summarized 
in Table~\ref{tab:dens_mass}.  
The total mass in each region, $\sim$0.01\,M$_\odot$, is consistent with 
our previous estimates \citep{Katsuda2008}, and is comparable with typical 
OFMKs in the remnant \citep{Winkler1988}.

\section{Discussion}

The {\it XMM-Newton} RGS allowed successful measurements of accurate
Doppler velocities of two X-ray--emitting ejecta features (i.e., the 
ejecta knot and the ejecta filament) in the NE quadrant of Puppis~A.  
In our spectral analysis, we divided each feature into four cross-dispersion 
regions.  The error-weighted mean Doppler velocities for the four regions are 
obtained to be 1480$\pm$140$\pm$60\kms\ blueward for the ejecta knot and 
650$\pm$70$\pm$60\kms\ redward for the ejecta filament, where the 
first and second term errors, respectively, represent the standard 
deviation for the four regions and a wavelengths accuracy of the 
RGS\footnote{http://xmm2.esac.esa.int/docs/documents/CAL-TN-0030.pdf}.
The Doppler velocity of the ejecta knot is fully consistent with 
previous optical measurements for the $\Omega$ filament 
\citep{Winkler1985}, ensuring robustness of our measurements.

The RGS spectra also enabled, for the first time, to measure broadening
of O (and Ne) K-shell lines of the ejecta knot.  The broadening is found to 
be modest, $\sigma\lesssim0.9$\,eV at \oviii\ Ly$\alpha$, indicating an 
upper limit of an oxygen temperature of 30\,keV.  As far as we know, 
this is the second result that we were able to measure an oxygen 
temperature of fast-moving ejecta knots in X-rays.  Comparing our results 
with the first successful object, a comparably fast-moving ejecta knot 
in the northwestern rim of SN~1006, we notice that the oxygen temperature 
in Puppis~A's knot is at least an order-of-magnitude lower than that in 
SN~1006's knot \citep[$kT_{\rm O}\sim$300\,keV:][]{Vink2003,Broersen2013}.

The post-shock temperature, $T_i$, for species $i$ with mass $m_i$ right 
behind a collisionless shock in the absence of the magnetic pressure is 
given as \citep[e.g.,][]{Spitzer1965}:
\begin{eqnarray}
\label{eq:init_temp}
kT_i = \frac{3}{16} m_i v_{\rm sh}^2
\end{eqnarray}
for shock velocity, $v_{\rm sh}$.  Thus, the particles will have 
mass-proportional temperatures in the limit of collisionless plasma.  
In fact, the high oxygen temperature together 
with a relatively low electron temperature of $\sim$1.5\,keV in SN~1006 
has been reasonably interpreted as evidence for temperature 
nonequilibration between ions and electrons after collisionless shock 
heating.  As for Puppis~A's knot, the oxygen temperature is expected to 
be $\sim$130\,keV, if we assume that the shock velocity is similar to 
the gas velocity: $\sim2000$\kms derived by the Doppler velocity of 
$\sim1500$\kms and the optical proper motion of $\sim$1250\kms\ at a 
distance of 2.2\,kpc \citep{Winkler1988,Garber2010,Reynoso2003}.  This 
prediction conflicts with our measurement, $kT_{\rm O}\lesssim$30\,keV.

To understand the apparent temperature inconsistency between the 
measurement and the simple prediction above, we are required to inspect 
heating processes at the collisionless shocks.  In fact, it is well known 
that temperatures measured in the immediate post-shock regions do not 
often obey Equation \ref{eq:init_temp} 
\citep[][for a recent review]{Ghavamian2007,Vink2012}.  
The deviation may be due to plasma processes that raise an electron 
temperature at (in front of) the shock 
\citep[][]{Ohira2008,Rakowski2008} and/or significant energy 
leakage into cosmic rays that lowers post-shock temperatures of all 
particle species \citep[][]{Hughes2000,Helder2009}.  In any case, 
oxygen can have a lower temperature than that predicted by 
Equation \ref{eq:init_temp}.

\begin{figure}
\includegraphics[angle=0,scale=0.6]{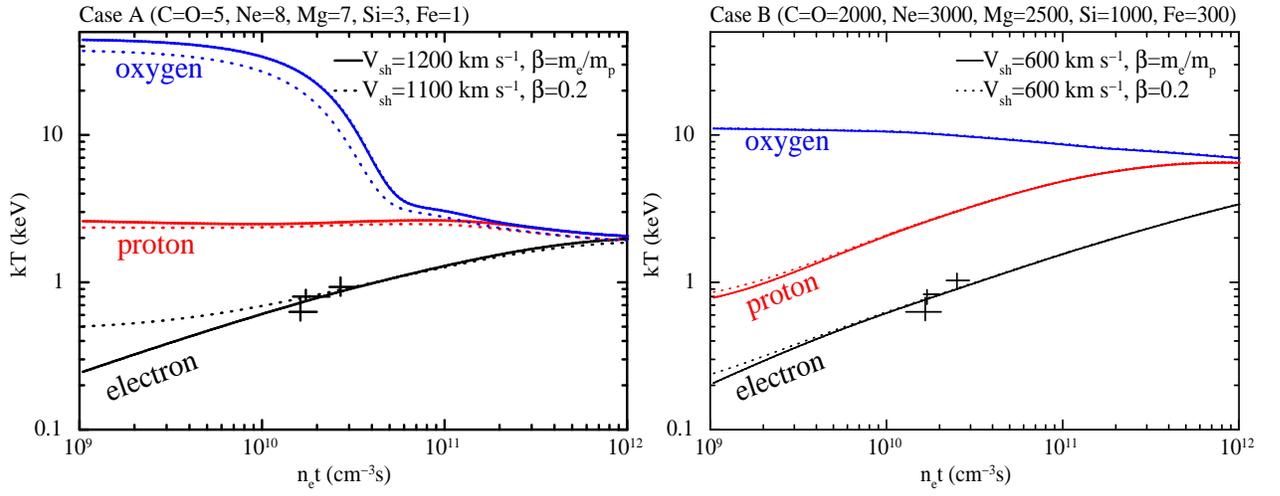}
\caption{Temperature histories for electrons (black), protons 
  (red), and oxygen (blue) as a function of $n_{\rm e}t$.
  The models are calculated for two initial electron-to-proton 
  temperature ratios, $\beta$, of the mass ratio (solid curve) and 
  0.2 (dotted curve).  The crosses are electron temperatures and 
  ionization timescales measured for the three regions in the ejecta 
  knot (see, Table~\ref{tab:MOSresults}).  
  Left and right panels are calculated for case A (several-solar 
  abundances as indicated on top of the plot) and case B 
  (super-solar abundances), respectively. 
} 
\label{fig:temp_equil}
\end{figure}

In this context, it is important to clarify whether the 
$T_{\mathrm O}$/$T_{\mathrm e}$ ratio observed in Puppis~A's knot can 
be explained by temperature equilibration due to Coulomb interactions 
behind a collisionless shock or requires other scenarios.
To this end, assuming pure collisionless shock heating, i.e., 
initial temperatures are given by Equation \ref{eq:init_temp}, 
we solve the following temperature equilibration due to Coulomb 
interactions \citep[e.g.,][]{Spitzer1965}:
\begin{eqnarray}
\label{eq:temp_equil}
\frac{dT_i}{dt} = \frac{T_j - T_i}{t_{\rm{eq}(i,j)}},~~
t_{\rm{eq}(i,j)} = 5.87\frac{A_i A_j}{n_j Z_i^2 Z_j^2 \rm{ln(\Lambda)}}
\left( \frac{T_i}{A_i} + \frac{T_j}{A_j} \right)^\frac{3}{2}
\end{eqnarray}
for species $i$ and $j$ with atomic numbers $A_i$ and $A_j$, charges 
$Z_i$ and $Z_j$, number density $n_j$, and the Coulomb logarithm 
ln($\Lambda$) $\sim$ 40.  We coupled the differential equations for 
electrons, protons, He, C, O, Ne, Mg, Si, and Fe, using the abundances 
for either case A (He/H=1, C/H=O/H=5, Ne/H=8, Mg/H=7, Si/H=3, and Fe/H=1 
times the solar values, see Table~\ref{tab:MOSresults}) or case B 
(He/H=1, C/H=O/H=2000, Ne/H=3000, Mg/H=2500, Si/H=1000, and Fe/H=300 
times the solar values).  After examining several shock velocities, we 
find that the electron temperatures and ionization timescales in the 
ejecta knot can be best explained at $v_{\rm sh}\sim1200$\kms\ or 
$v_{\rm sh}\sim600$\kms\ for case A or case B, respectively.  
The temperature histories for 
electrons, protons, and oxygen as a function of \n_et\ are shown in 
Figure~\ref{fig:temp_equil}.  From the plot, we also find that the oxygen 
temperature measured, $T_{\rm O}\lesssim30$\,keV, is consistent with 
the expectation in both cases.  Therefore, we do not need to invoke 
some exotic scenarios to explain the low $T_{\rm O}/T_{\rm e}$ ratio.

Still, the temperatures measured may be also reproduced for the case 
that initial post-shock temperatures are not proportional to particle 
masses, as is often observed.  If it is the case, the shock velocities would 
be different, causing uncertainties on the shock velocity.  Recent 
observational studies agree in that the initial electron-to-proton 
temperature ratio, $T_{\rm e}/T_{\rm p} = \beta$, is close to unity 
(full equilibration) for slow shocks ($v_{\rm sh}\lesssim$400\kms), 
but gradually goes down for faster shocks 
\citep[][and references therein]{Ghavamian2007,Heng2010}.  
We take the value of $\beta$ to be 0.2 which is typical around 
$v_{\rm sh} = 1000$\kms\ \citep{vanAdelsberg2008}.  We find 
that temperature histories expected at $v_{\rm sh}\sim1000$\kms\ (case A)
or $v_{\rm sh}\sim600$\kms\ (case B), indicated as dotted lines in 
Figure~\ref{fig:temp_equil}, best match our measurements.  
In this context, we argue that the shock velocity does not depend on the 
degree of initial temperature equilibration compared with the 
ambiguity of abundances.

It is interesting to note that the inferred shock velocity, 600--1200\kms, 
is less than the shocked-gas velocity, $\sim2000$\kms.  This suggests that 
the ejecta knot was heated by a reverse shock rather than a forward shock,
since the forward shock velocity should be $\sim$2500\kms according to the 
Rankine-Hugoniot relation: 
\begin{eqnarray}
\label{eq:fs_rankine}
v_{\rm fs} = \frac{\gamma + 1}{2}v_{\rm sg},
\end{eqnarray}
where $v_{\rm fs}$ and $v_{\rm sg}$ are velocities of the 
forward shock and the shocked gas, respectively, and $\gamma$ is 
the specific heat ratio taken to be 5/3 for a non-relativistic ideal gas.  
The Rankine-Hugoniot relation over the reverse shock can be written 
down as:
\begin{eqnarray}
\label{eq:rs_rankine}
v_{\rm free} - v_{\rm rs} = \frac{\gamma + 1}{\gamma - 1}
\left(v_{\rm sg} - v_{\rm rs}\right),
\end{eqnarray}
where $v_{\rm free}$ and $v_{\rm rs}$ are velocities of the 
freely-expanding ejecta and the reverse shock in the observer's 
frame.  Substituting $v_{\rm free} - v_{\rm rs}$ (=$v_{\rm sh}$) 
= 600--1200\kms, $v_{\rm sg}=$ 2000\kms, and $\gamma=5/3$, we 
compute the value of $v_{\rm free}$ to be 2900--2450\kms, depending on 
the shock velocity of 600--1200\kms.  This velocity agrees with a 
typical ejecta velocity in O-rich layer for Type IIb SN 
\citep{Houck1996,Iwamoto1997} which is likely the origin of Puppis~A 
\citep{Chevalier2005}.  Also, the distance, 0.3--0.6\,pc, that the 
shock of this velocity travels for $\sim$500\,years (which is a product 
of the ionization timescale, 3$\times10^{10}$\cm3s, divided by the 
electron density, 2\,cm$^{-3}$, as in Tables~\ref{tab:MOSresults} and 
\ref{tab:dens_mass}) is comparable with the size of the knot 
($\sim$0.6\,pc), showing self-consistency of our interpretation.

We comment on the remarkable difference in $T_{\rm O}/T_{\rm e}$ between 
Puppis~A's knot and SN~1006's knot.  The proper motion of SN~1006's 
knot is measured from recent X-ray observations to be $\sim$3000\kms\ 
\citep{Katsuda2013}.  Provided that it is shocked by a forward shock, 
then the shock velocity would be the proper motion itself.  
On the other hand, if it is shocked by a reverse shock, the proper 
motion should be the velocity of the shocked gas.  Also, freely-expanding 
ejecta velocities is likely $\sim$10000\kms\ \citep[][]{Iwamoto1999}, 
since there is a general 
consensus that SN~1006 is a Type Ia SNR.  This is indeed comparable to a 
freely-expanding Si-rich ejecta velocity of $\sim$7000\kms\ measured by 
the UV absorption edge \citep{Hamilton2007} as well as a simple 
estimate from the radius at the knot ($\sim$9\,pc) and the SNR age 
($\sim$1000\,yr).  Then, Equation~\ref{eq:rs_rankine} gives the reverse 
shock velocity in the ejecta-rest frame to be $\sim$9300\kms (or 
$\sim$5300\kms, if we take $v_{\rm free}$ to be 7000\kms).
Therefore, the shock velocities estimated for SN~1006's knot are at 
least a few times faster than that for Puppis~A' knot.  
In addition, the ionization timescale is an order-of-magnitude lower 
in SN~1006's knot than in Puppis~A' knot.  These differences can 
qualitatively explain the different oxygen temperatures.

Future high-resolution X-ray spectroscopy with the nondispersive soft 
X-ray spectrometer onboard Astro-H 
\citep[FWHM$\sim$5\,eV:][]{Takahashi2010,Mitsuda2010} 
will reveal ion and electron temperatures for many SNRs.  Hopefully, 
we will be able to measure ion temperatures in different ionization 
states for every species.  Such information will significantly improve 
our understanding on temperature equilibration for fast collisionless 
shocks.  One particularly interesting target in Puppis~A would be the 
Si-rich clump in the NE remnant \citep{Hwang2008}, for which the RGS 
cannot produce high-resolution 
spectra due to its large angular size (3$^{\prime}\times5^{\prime}$).  
Fortunately, this target is scheduled to be observed in 2013 January 
with a soft X-ray spectrometer onboard the Micro-X sounding rocket 
experiment\footnote{http://space.mit.edu/micro-x/index.html} 
\citep[FWHM$\sim$2\,eV:][]{Figueroa-Feliciano2012}.  This experiment
will provide exciting results shortly.

\section{Summary}

We have presented results from our new {\it XMM-Newton} RGS and MOS 
observations of two ejecta features in the Galactic SNR Puppis~A: one is
an ejecta knot positionally coincident with the optical $\Omega$ filament 
\citep{Winkler1985}, and the other is an ejecta filament located near the 
NE edge of the remnant.  The results obtained are summarized below:

\begin{itemize}

\item The Doppler velocity of the ejecta knot is measured to be 
1480$\pm$140$\pm$60\kms\ blueward, which is fully consistent with 
previous optical measurements for the $\Omega$ filament \citep{Winkler1985}.  
We find that the Doppler velocities are constant within the knot, except 
for a northernmost piece where a slightly slower velocity is indicated.

\item An upper limit of line broadening for O K-shell lines in
the ejecta knot is obtained to be $\sim$0.9\,eV, leading to an oxygen 
temperature 
of $\lesssim30$\,keV.  This upper limit, combined with the electron 
temperature and the ionization timescale derived from the MOS 
spectra, suggests that the ejecta knot was heated by a reverse 
shock whose velocity is $\sim$600--1200\kms, without requiring 
rapid temperature equilibration due to plasma instabilities nor 
temperature dropping due to energy leakage into cosmic rays.

\item The oxygen temperature of the ejecta knot is an order-of-magnitude 
lower than that obtained in a comparably fast-moving ejecta knot in 
SN~1006 \citep{Vink2003,Broersen2013}.  We argue that the difference
would be due to both different shock velocities and the ionization 
timescale.

\item The Doppler velocity for the ejecta filament is also measured, 
for the first time, to be 650$\pm$70$\pm$60\kms\ redward.

\end{itemize}

\acknowledgments

We thank the referee, Knox Long, for a number of comments that help 
clarify the paper.  S.K.\ is supported by the Special Postdoctoral 
Researchers Program in RIKEN.  This work is partly supported by a 
Grant-in-Aid for Scientific Research by the Ministry of Education, 
Culture, Sports, Science and Technology (23-000004 and 24-8344).


\begin{deluxetable}{llcccccccc}
\tabletypesize{\tiny}
\tablecaption{Spectral-fit paramters derived from the RGS}
\tablewidth{0pt}
\tablehead{
{Component} & {Parameter} &\multicolumn{4}{c}{Regions}}
\startdata
& & A & B & C & D \\\hline
Knot & Redshift (10$^{-3}$) & -4.5$^{+0.3}_{-0.1}$, -4.1$^{+0.1}_{-0.5}$ & -5.0$\pm$0.2, -4.6$\pm$0.2 & -5.3$^{+0.3}_{-0.1}$, -5.3$^{+0.3}_{-0.1}$ & -5.2$\pm$0.2 \\
& Line broadening ($\sigma$ in eV @0.654\,keV) & 0.0 ($<0.4$), 0.0 ($<0.7$) & 0.3$^{+0.4}_{-0.2}$, 0.9$\pm$0.3 & 0.0 ($<0.4$), 0.0 ($<0.4$) & 0.8$\pm$0.3 \\
& Intensities:~~O He$\alpha$ (f)  & 411$\pm$101 & 812$\pm$123 & 882$\pm$126 & 657$\pm$104 \\
& ~~~~~~~~~~~~~~~~O He$\alpha$ (i)  & 6 ($<87$) & 0 ($<49$) & 113$\pm$104 & 0 ($<74$) \\
& ~~~~~~~~~~~~~~~~O He$\alpha$ (r)  & 702$\pm$103 & 1623$\pm$133 & 1519$\pm$127 & 1085$\pm$116 \\
& ~~~~~~~~~~~~~~~~O He$\beta$  & 157$\pm$30 & 342$\pm$38 & 428$\pm$40 & 134$\pm$33 \\
& ~~~~~~~~~~~~~~~~O Ly$\alpha$  & 956$\pm$54 & 1898$\pm$64 & 863$\pm$52 & 220$\pm$37 \\
& ~~~~~~~~~~~~~~~~O Ly$\beta$  & 123$\pm$20 & 247$\pm$24 & 107$\pm$21 & 40$\pm$16 \\
& ~~~~~~~~~~~~~~~~Ne He$\alpha$ (f)  & 26$\pm$25 & 115$\pm$29 & 139$\pm$28 & 48$\pm$22 \\
& ~~~~~~~~~~~~~~~~Ne He$\alpha$ (i)  & 0 ($<6$) & 25 ($<54$) & 42$\pm$28 & 9 ($<30$) \\
& ~~~~~~~~~~~~~~~~Ne He$\alpha$ (r)  & 149$\pm$25 & 216$\pm$29 & 191$\pm$28 & 47$\pm$22 \\
\hline 
Filament & Redshift (10$^{-3}$) & 2.3$^{+0.2}_{-0.5}$ & 1.9$^{+0.4}_{-0.1}$ & 2.2$^{+0.1}_{-0.3}$ &  2.4$^{+0.4}_{-0.2}$ \\
& Intensities:~~O He$\alpha$ (f)  & 285$\pm$112 & 489$\pm$118 & 374$\pm$123 & 207$\pm$179 \\
& ~~~~~~~~~~~~~~~~O He$\alpha$ (i)  & 0 ($<48$) & 77 ($<185$) & 70 ($<167$) & 129$\pm$104 \\
& ~~~~~~~~~~~~~~~~O He$\alpha$ (r)  & 459$\pm$101 & 612$\pm$109 & 1069$\pm$119 & 619$\pm$117 \\
& ~~~~~~~~~~~~~~~~O He$\beta$  & 26 ($<63$) & 28 ($<69$) & 213$\pm$38 & 366$\pm$43 \\
& ~~~~~~~~~~~~~~~~O Ly$\alpha$  & 237$\pm$41 & 167$\pm$40 & 210$\pm$38 & 223$\pm$40 \\
& ~~~~~~~~~~~~~~~~O Ly$\beta$  & 113$\pm$35 & 56$\pm$34 & 0 ($<34$) & 0 ($<7$) \\
\hline 
Diffuse & $kT_{\mathrm e}$ (keV) & 0.46$\pm$0.01 & 0.47$\pm$0.01 & 0.44$\pm$0.01 & 0.45$\pm$0.01 \\
\hline 
$\chi^{2}$/d.o.f. & & 1937.8 / 1260 & 2073.0 / 1280 & 1841.6 / 1260 & 1875.9 / 1257 \\
\enddata
\label{tab:RGSresults}
\tablecomments{$N_{\mathrm H}$ values are fixed to 3$\times10^{21}$\,cm$^{-2}$.  Line intensities are in units of 10$^{-5}$ photons\,cm$^{-2}$\,s$^{-1}$, respectively.  Two values of the redshifts and line broadening in Regions A, B, and C are derived by the RGS1 (left) and the RGS2 (right).  The errors represent 90\% confidence levels on an interesting single parameter.}
\end{deluxetable}

\begin{deluxetable}{lcccccc}
\tabletypesize{\tiny}
\tablecaption{Spectral-fit parameters derived from the MOS}
\tablewidth{0pt}
\tablehead{
\colhead{Parameter}&\multicolumn{2}{c}{North}&\multicolumn{2}{c}{West}&\multicolumn{2}{c}{South}}
\startdata
& Case A & Case B & Case A & Case B & Case A & Case B \\
\hline
$kT_{\mathrm e}$ (keV) & 0.93$^{+0.15}_{-0.12}$ & 1.03$\pm$0.12 & 0.80$^{+0.09}_{-0.05}$ & 0.83$^{+0.06}_{-0.12}$ & 0.63$^{+0.13}_{-0.08}$ & 0.63$^{+0.16}_{-0.08}$\\
log($n_{\mathrm e}t$/cm$^{-3}\,$s) & 10.43$^{+0.09}_{-0.06}$ & 10.40$^{+0.07}_{-0.06}$ & 10.24$^{+0.13}_{-0.07}$ & 10.23$^{+0.07}_{-0.02}$ & 10.21$^{+0.09}_{-0.06}$& 10.22$^{+0.09}_{-0.11}$\\
(C/H)/(C/H)$_\odot$ & 11.9$^{+8.0}_{-7.0}$ & 6000$^{+2400}_{-3800}$ & 5.6$^{+6.2}_{-5.6}$ & 1800$\pm$1800 & 3.1$^{+3.6}_{-2.9}$ & 1500$\pm$1300 \\
(N/H)/(N/H)$_\odot$ & 0.0 ($<1.2$) & 0 ($<300$) & 1.8 ($<3.8$) & 500 ($<1100$) & 1.5$^{+1.4}_{-1.1}$ & 700$\pm$500 \\
(O/H)/(O/H)$_\odot$ & 5.2$^{+1.1}_{-0.8}$ & 2000$^{a}$ & 6.9$^{+1.1}_{-0.8}$ & 2000$^{a}$ & 4.5$^{+1.1}_{-0.8}$ & 2000$^{a}$ \\
(Ne/H)/(Ne/H)$_\odot$ & 6.5$^{+1.4}_{-1.1}$ & 2400$^{+100}_{-200}$ & 10.3$^{+2.4}_{-2}$ & 3000$\pm$200 & 7.5$^{+2.2}_{-1.7}$ & 3400$\pm$300 \\
(Mg/H)/(Mg/H)$_\odot$ & 5.6$^{+1}_{-0.9}$ & 2100$\pm$200 & 7.9$^{+0.7}_{-1.2}$ & 2300$\pm$200 & 7.2$^{+1.9}_{-1.5}$ & 3200$\pm$500 \\
(Si/H)/(Si/H)$_\odot$ & 3.2$\pm$0.7 & 1200$^{+200}_{-100}$ & 2.8$^{+1.1}_{-1.2}$ & 900$\pm$300 & 4.0$^{+1.9}_{-1.4}$ & 1900$^{+600}_{-500}$ \\
(Fe/H)/(Fe/H)$_\odot$ & 1$^{a}$ & 380$^{+70}_{-60}$ & 1$^{a}$ & 270$^{+90}_{-110}$ & 1$^{a}$ & 390$^{+130}_{-110}$ \\
Redshift ($10^{-3}$) & -6.5$^{+0.6}_{-0.3}$& -6.6$^{+1.0}_{-0.1}$ & -6.1$^{+0.3}_{-1.0}$& -6.1$^{+0.3}_{-1.1}$ & -5.4$^{+1.6}_{-0.2}$& -5.4$^{+1.0}_{-0.1}$\\
$n_{\mathrm e}n_{\mathrm H}V/4\pi d^{2}$ (10$^{7}$cm$^{-5}$) & 740$^{+130}_{-140}$ & 1.84$^{+0.17}_{-0.19}$ & 240$^{+60}_{-10}$ & 0.80$^{+0.07}_{-0.04}$ & 460$^{+170}_{-160}$& 1.04$^{+0.15}_{-0.18}$\\
\hline 
$\chi^{2}$/d.o.f. & 187.7 / 126 & 200.6 / 126 & 141.0 / 103 & 148.3 / 103 & 180.8 / 108 & 192.1 / 108 \\
\enddata
\label{tab:MOSresults}
\tablecomments{$^a$(Fe/H)/(Fe/H)$_\odot$ and (O/H)/(O/H)$_\odot$ are fixed to 1 (Case A) and 2000 (Case B), respectively.  $N_{\mathrm H}$ values are fixed to 3$\times10^{21}$\,cm$^{-2}$. 
The errors represent 90\% confidence levels.}
\end{deluxetable}

\begin{deluxetable}{lcccccc}
\tabletypesize{\tiny}
\tablecaption{Densities (10$^{-4}$\,cm$^{-3}$) and Masses (10$^{-4}$\,M$_\odot$) in the ejecta knot}
\tablewidth{0pt}
\tablehead{
\colhead{Parameter}&\multicolumn{2}{c}{North}&\multicolumn{2}{c}{West}&\multicolumn{2}{c}{South}}
\startdata
& Case A & Case B & Case A & Case B & Case A & Case B \\
\hline
$n_{\rm e}$ & 44000 & 11000 & 33000 & 7800 & 37000 & 7200\\
$n_{\rm H}$ & 35000 & 360 & 26000 & 370 & 29000 & 340 \\
$n_{\rm C}$ & 150 & 770 & 53 & 240 & 33 & 190\\
$n_{\rm O}$ & 150 & 610 & 150 & 620 & 110 & 570\\
$n_{\rm Ne}$ & 28 & 110 & 33 & 130 & 27 & 140\\
$n_{\rm Mg}$ & 7.5 & 28 & 7.7 & 31 & 8.1 & 41\\
$n_{\rm Si}$ & 4.0 & 15 & 2.6 & 11 & 4.2 & 23\\
$n_{\rm Fe}$ & 1.6 & 6.3 & 1.2 & 4.7 & 1.4 & 6.2\\
\hline
$M_{\rm H}$ & 82 & 0.8 & 36 & 0.5 & 62 & 0.7\\
$M_{\rm C}$ & 4.3 & 22 & 14 & 4.0 & 0.8 & 4.7\\
$M_{\rm O}$ & 5.8 & 23 & 3.3 & 14 & 3.8 & 19\\
$M_{\rm Ne}$ & 1.3 & 5.0 & 0.9 & 3.7 & 1.1 & 5.9\\
$M_{\rm Mg}$ & 0.4 & 1.6 & 0.3 & 1.0 & 0.4 & 2.1\\
$M_{\rm Si}$ & 0.3 & 1.0 & 0.1 & 0.4 & 0.2 & 1.3\\
$M_{\rm Fe}$ & 0.2 & 0.8 & 0.1 & 0.4 & 0.2 & 0.7\\
\hline 
\enddata
\label{tab:dens_mass}
\tablecomments{Typical errors which are mainly due to the assumption of 
the plasma depth are about a factor of 2.}
\end{deluxetable}

\end{document}